# ADAPTING THE DEVS KERNEL 'RT-CADMIUM' TO THE ESP32 EMBEDDED PLATFORM

SASISEKHAR MANGALAM GOVIND

JOHN SAHAYA RANI ALEX

GABRIEL WAINER

**ABSTRACT**

Discrete Event Modelling of Embedded Systems (DEMES) is a development methodology based on the Discrete Event Systems (DEVS) specification that improves the time -to-market by simplifying the development and testing of embedded systems. CADMIUM is a C++ header-only library developed at Carleton University that helps simulate models built using the DEVS specification. RT-CADMIUM is a fork of CADMIUM that provides a development framework that helps users develop systems using the DEMES technology. RT-CADMIUM, however, has a limited scope of deployment due to the use of Mbed OS as its Hardware Abstraction Layer (HAL). This paper provides the methodology for porting the RT-CADMIUM library to a different platform (ESP32 specifically). This paper also portrays the performance improvements gained due to this adoption.

**INTRODUCTION**

In the current era of rapid technological advancement, efficient software design is of utmost importance, as is the optimization of firmware and associated hardware development. The proliferation of 'Smart' appliances has increased the need for IoT-enabled embedded systems[1]. Real-time systems are a vital category of IoT systems that require the timely processing and communication of data to achieve the desired system behaviour and meet the application requirements[2]. They facilitate real-time data acquisition, processing, and the implementation of complex synchronous communication protocols, among other things. A real-time system comprises interconnected subsystems or components that interact with the environment in response to real-time stimuli, resulting in immediate system responses. As a result, stringent timing requirements must be met to achieve such instantaneous response times[3]. However, developing a real-time controller is a challenging task, both technically and financially[4].

Discrete Event Modelling of Embedded Systems (DEMES) is a development methodology based on the Discrete Event System (DEVS) specification[5]. Systems modelling and simulation tend to be widely used in the early stages of development of projects but tend to be abandoned as the project moves from paper to the real world. But maintaining a model for a given system throughout the development cycle would allow for simulation to test the reliability and robustness of the system in conditions that maybe impractical to replicate physically. Since DEMES is based on the formal modelling and simulation paradigm DEVS, it allows the developer to maintain the system models throughout the development cycle[5]. By developing a kernel capable of executing models created using the DEVS formalism, it becomes possible to simulate theoretical systems designed by researchers on a microcontroller, thereby realizing them in the physical world[6]. DEVS specification revolves around timed events, hence the development of a kernel that executes said models would be able to follow very hard timing constraints. The modularity of a model/ system designed using the DEVS specification allows for any module (called atomic models) to be replaced/ upgraded without affecting the operation of the complete system (called coupled model). This would also imply that once a generic atomic model is created, it can be used in other systems that require its functionality, hence reducing the time to market exponentially.

There exist various DEVS simulators, such as XDEVS from the University of Barcelona and PowerDEVS from the University of Buenos Aires. However, this paper will focus on CADMIUM, which was developed at Carleton University. CADMIUM is a C++ header-only library that enables users to model and simulate DEVS models[7]. RT-CADMIUM, a fork of CADMIUM, permits users to develop DEVS models, simulate them, and execute them in real-time on ARM microcontrollers that support the Mbed OS platform[8]. The principal aim of this research paper is to enhance the RT-CADMIUM kernel. The proposed enhancements involve two main aspects: Firstly, upgrading the foundation of RT-CADMIUM to the latest version of CADMIUM[7] developed by Roman Cardenas. Secondly, this paper proposes a methodology to adapt RT-CADMIUM to other platforms while eliminating its dependence on Mbed OS. The study also intends to enhance the performance of RT-CADMIUM to render it more feasible for commercial deployment. Additionally, the research aims to refine the software-hardware co-design aspects of the development framework.

**RELATED WORK**

DEVS (Discrete Event System) specification is a widely used formalism for modelling complex dynamic systems using discrete-event abstraction created by University of Arizona Prof. Bernard P. Ziegler. One of the reasons DEVS was chosen as the basis of the kernel is that the formalism defines both the system structure and the system behaviour[6]. DEVS consists of two types of models: atomic models and coupled models. Atomic models represent the behaviour of individual components, including their current state, the time they will remain in that state, and their input and output ports. On the other hand, coupled models are used to connect groups of models, whether atomic or coupled, by passing outputs from one DEVS model to the inputs of another. These links are created using the coupled model, which can contain both types of DEVS models. Coupled models are useful for creating modular hierarchical designs, which can be easily adapted and modified as necessary[1]. The input/ output events along with the states define the behaviour of a system based on DEVS. The 7-tuple definition of a DEVS atomic model:

$$M = <X, Y, S, s_0, ta, \delta_{int}, \delta_{ext}, \delta_{con}, \lambda>$$

- X is the set of inputs.
- Y is the set of outputs.
- S is the set of states
- $s_0$ is an individual state ($s_0 \in S$)
- $ta: S \to \mathbb{R}_{>0} \cup \infty$ is the time advance function.
- $\delta_{int}: S \to S$ is the internal transition function. The model transitions from $s_0 \in S$ to $s_1 \in S$ after spending $ta(s_0)$ time in state $s_0$ without receiving an input.
- $\delta_{ext}: S \times \mathbb{R} \times X \to S$ is the external transition function. This function is triggered when a set of inputs $X_b \subseteq X$ after $e$ has elapsed since the model entered state $s \in S$.
- $\delta_{con}: S \times \mathbb{R} \times X \to S$ is the confluent transition function and is responsible for ensuring collisions don't occur. This function is triggered when an internal and an external transition occur at the same time. Generally, the internal transition would be executed first. This function is the main difference between Parallel DEVS (used for RT-Cadmium) and Classic DEVS.
- $\lambda: S \to Y$ is the output function. Triggered right before the internal transition from $s_0 \in S$ to $s_1 \in S$, and generates outputs $\lambda(s_0) = Y_b \subseteq Y$ for state $s_0$.

The formal structure of a coupled DEVS modul 8-tuple representation:

$$N = <X, Y, C, EIC, EOC, IC>$$

- X is the set of input events
- Y is the set of output events

- C is the set of submodels. Any element $c \in C$, is either an atomic or a coupled model defined inside the coupled model.
- EIC is the External Input Coupling. Defines the connections from models outside N to the components $c \in C$.
- EOC is the External Output Coupling. Defines the connections from the components within N to models outside N.
- IC defines the connections between any component $c_i \in C$ and $c_j \in C$.

The coupled model carries forward the property of any DEVS model. This closure under coupling [9], allows for construction of hierarchical models, whose behaviour remain consistent and predictable[6]. Further, the coupled model defines the structure of the complete system. DEVS decouples model, experiments, and execution engines (allowing for portability and interoperability)[5].

CADMIUM is a tool developed by Carleton university that helps simulate DEVS based models. RT-CADMIUM is a real-time kernel based on CADMIUM that allows for these DEVS models to be implemented in hardware[1]. The interface between the root co-ordinator and the clock of the platform is what enables the CADMIUM to execute the models in real time. The Algorithm that enables the same is shown in Figure 1.

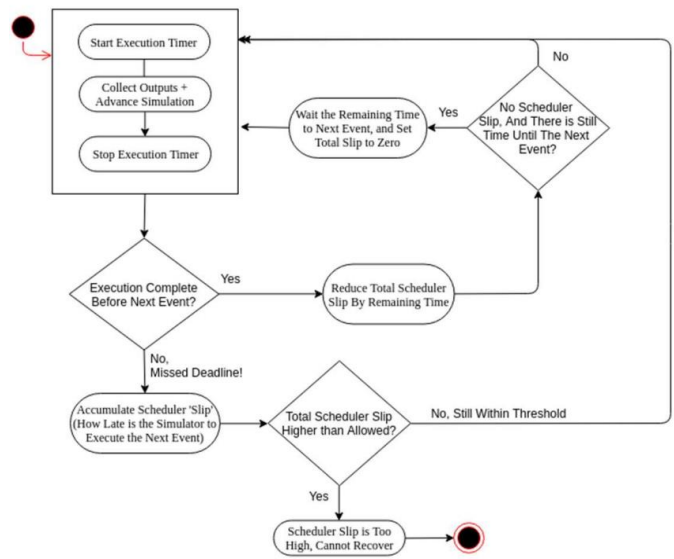

*Figure 1: Real-time clock with Scheduler slip adjustment* [1]

The two main jobs of the co-ordinator are to 1) Collect the outputs 2) Advance the simulation. After these tasks are completed, the simulation moves to the next state. Hence, the simplest timer will advance the collect the outputs, advance the simulation, and then wait for the next event. This ideal time scheduler does not consider the time taken by the platform to evaluate the model, collect the outputs and advance the simulation. This non-zero time would disrupt the simple timer algorithm. Hence, the algorithm shown in Figure 1 was developed over multiple iterations and handles scheduler slip. A user configurable flag can be modified to allow a certain degree of time slip to be permissible. This algorithm employs two timers, execution timer and wait (referred to as timeout timer in the actual code) timer. The execution timer starts at the beginning of an event, runs while the engine collects the output, and stops after the simulation has advanced. Once the simulation has advanced, the algorithm checks if the execution was completed before the start of the next event. If the execution took longer than the assigned deadline, the user is notified about the missed deadline, and a variable is defined to keep track of this slip; the total slip is accumulated over time. If the accumulated value exceeds the scheduler slip allowance, the algorithm halts program execution; else, the algorithm starts the execution of the next event. If, however, the deadline was not missed, the time left to the next event is subtracted from the accumulated slip value. Moreover, if the delta between the current time and the start time of the next event is non-zero, indicating the previous event completed execution prematurely, a wait timer is employed to stall the execution till the start of the next event[1].

This algorithm is implemented in the "rt_clock.hpp" file within the library.

**RT-CADMIUM: IMPLEMENTATION**

The system under discussion employs a generic C++ approach for kernel operations and scheduling, which allows for flexibility and portability across different hardware platforms. However, the implementation of the "rt_clock.hpp" file uses API calls to the Mbed OS hardware abstraction layer (HAL)[1].

This reliance on Mbed OS API calls limit the portability of the library to only those microcontrollers and embedded boards that support Mbed OS. The HAL provides a standardized interface for working with hardware, but it also introduces dependencies that restrict the system's compatibility with other platforms[10].

"rt_clock.hpp", as mentioned earlier, contains 2 timers: the execution timer and the timeout timer. The execution timer tracks the simulation time, which is the elapsed time from the start of the simulation. This timer provides valuable information for identifying bottlenecks and issues in the system's performance. By monitoring the execution time, the system can optimize its performance and ensure efficient use of resources[1].

The timeout timer, on the other hand, tracks the time between internal transitions called sigma. In this context, sigma refers to the time between two events within the system (returned by the time advance function mentioned earlier), such as the arrival of new data or the completion of a task. The timeout timer ensures that the system remains responsive and prevents resource overuse by setting appropriate timeout values. The use of Mbed OS provides a standardized interface for working with hardware, but it also introduces dependencies that restrict the system's compatibility with other platforms[1].

Despite this limitation, the use of Mbed OS can still provide benefits in terms of simplicity and ease of use, particularly for developers who are familiar with the Mbed OS API. However, if portability across a wider range of platforms is a priority, alternative approaches that avoid platform-specific dependencies should be considered.

**RT-CADMIUM: METHODOLOGY OF ADAPTATION**

The software/ firmware that directly interacts between the user code and the hardware communication network of the underlying SoC can be regarded as the Hardware Abstraction Layer (HAL). The HAL provides APIs and function calls for the user to make use of, to expedite the development process. The standardization of HALs would allow for total software reusability and would help unify various development workflows[11]. However, it is worth noting that many microcontroller manufacturers provide HAL libraries as part of their development tools and support resources. Some examples of microcontroller manufacturers that offer HAL libraries include STMicroelectronics, Texas Instruments, NXP, and Microchip[12]–[15]. These proprietary HALs reduce the time of firmware development on microcontroller units (MCUs) manufactured by the respective companies, but HALs like Mbed OS aim to provide a wider HAL to support a multitude of microcontrollers from various manufacturers. Mbed OS is a monolithic kernel written in C and C++. It was (and continues) to be developed by ARM for low constrained devices[16]. Although Mbed OS promises a lower memory footprint, the lack of support for a major IoT platform like the ESP32 steers us away from this HAL. Replacement or removal of a HAL like Mbed OS would require an alternative HAL to be implemented in its stead. The target platform of interest is the ESP32. The ESP32, more specifically the ESP32-WROOM32D, is a powerful microcontroller that combines Wi-Fi (802.11 b/g/n), dual mode Bluetooth 4.2, and a multitude of peripherals including I2S, I2C, SPI, UART, etc. into a single piece of silicon[17], [18]. The SoC within, is based on two Xtensa LX6 microcontrollers that runs at 240MHz with 520kB of SRAM and upto 64 MB of flash storage[17]. Due to the low price, high performance, and flexible footprint, the ESP32 is a highly favourable option for developers to deploy an IoT system[19]. The ESP-IDF integrated development framework developed by Espressif, is a set of libraries designed in C that act as a HAL and provides developers with complete access to the hardware[18].

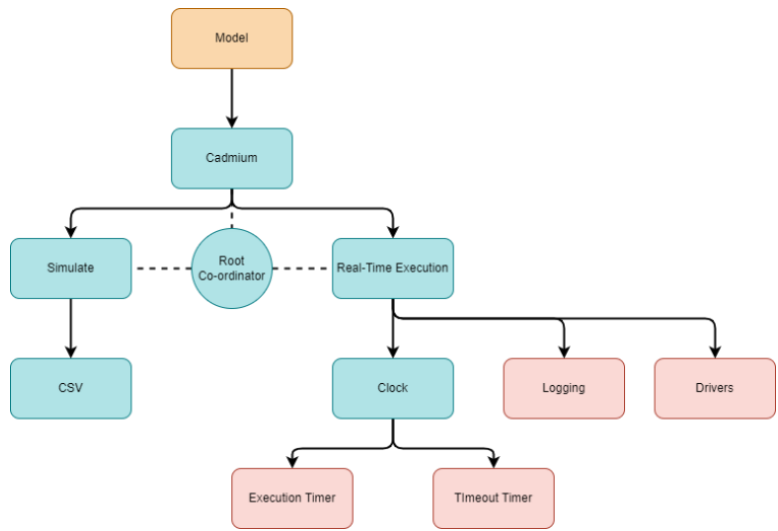

*Figure 2: Architecture of CADMIUM*

Figure 2: Architecture of CADMIUM shows the condensed architecture of CADMIUM. Once the model is fed into CADMIUM, the root co-ordinator handles the task of advancing the simulation. Depending on the platform of execution and the flags used during compilation, CADMIUM either simulates the model, or executes it on hardware (be it an embedded system running on RTOS, or a CISC system running Linux). 11 blocks can be seen to be a part of the diagram. Out of the 11 blocks, 4 are highlighted in red; they are: Execution timer, Timeout timer, Logging and Drivers. All the blocks are implemented in C++17 standard, but the highlighted blocks are unique in that they require implementation through hardware API calls (via the HAL). Each of the 4 blocks have a particular function:

Drivers:

The numerous files and pieces of code required to make it easier to integrate sensors and actuators from the physical layer into the CADMIUM layer are all represented comprehensively by the driver block. Specifically, this entails the inclusion of code that invokes APIs from HAL or other libraries, which encapsulate the interactions with the physical environment into Atomic DEVS models. The 2 main driver blocks are:

> Digital Input: The Digital Input atomic polls a given hardware pin with a predefined polling rate $\sigma$ and brings the Boolean data to the CADMIUM layer. For this atomic, the set of inputs $X = \varphi$ and set of output Y contains a singular Boolean port.

> Digital Output: The Digital Output atomic takes a Boolean input and reciprocates the same on a hardware pin. It moves the data from CADMIUM layer to the hardware layer. The set of inputs X contains a singular Boolean port, while the set of outputs $Y = \varphi$.

Logging:

This block represents the set of programs that allow the DEVS logging engine within CADMIUM to interact with the Serial Monitor (or any other forms of text display). Here, virtual functions for 'print to screen' are overridden using the appropriate APIs to interact with the output stream buffers.

Execution timer & Timeout timer:

As has already been discussed in the 'RELATED WORK' section, there are 2 timers that help in scheduling process. The real-time scheduling algorithm requires multiple API calls to the HAL to define the various timers, interrupt upon timer limit etc. The implementation of the same was done using the gptimers provided by the ESP-IDF.

**CASE STUDY: BLINKY**

A simple demonstration of the working and implementation of a system using CADMIUM is a Blinky program. A blinky is a simple system that allows us to observe the performance of the hardware running in conjunction with the CADMIUM kernel. The blinky system has an external input and an output enabling us to see the various response times of the system in different scenarios.

The blinky system has a simple task: Blink and LED with a period that switches between 2 values. What the period of oscillation would be for at any instance is controlled by a Boolean input into the system. The input toggles a flag. This flag is used as a switch to alternate between the 2 predefined values for the period. Finally, a Boolean output, which is connected to an LED, toggles based on the period. A model can be created that follows the DEVS specification, the block diagram of which is provided in Figure 3,

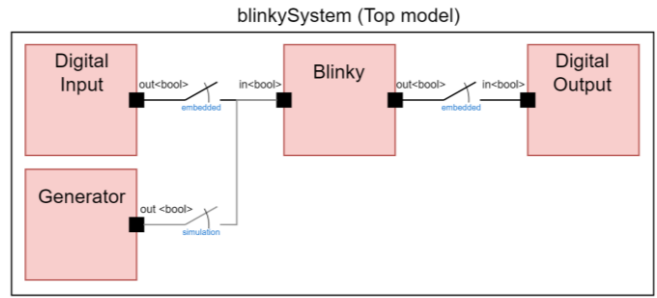

Figure 3: Block diagram of blinkySystem [20]

The blinkySystem model serves as the top-level model that contains all the atomics. The 'Digital Input' atomic is responsible for accepting user input and is compiled only when building for an embedded board and not during simulation. The Generator atomic model generates Boolean outputs randomly and is compiled only during simulation to provide inputs. The blinky atomic, which is compiled for both building for embedded boards and building for simulation, is responsible for receiving inputs, altering the oscillation period, and producing a Boolean output. Lastly, the Digital output block takes the input from the Blinky atomic and toggles the LED accordingly. This atomic is not compiled during simulation. The switches in Figure 3 aid in visualizing the cases in which the corresponding atomics are compiled or connected to the other atomics. When building for simulation, only the Generator and Blinky are compiled, and when building for deployment, the Digital Input, Blinky, and Digital Outputs are compiled. The implementation of the Digital Input block and Digital Output block have been displayed in previous sections. The implementation of the blinky model is represented by Figure 4.

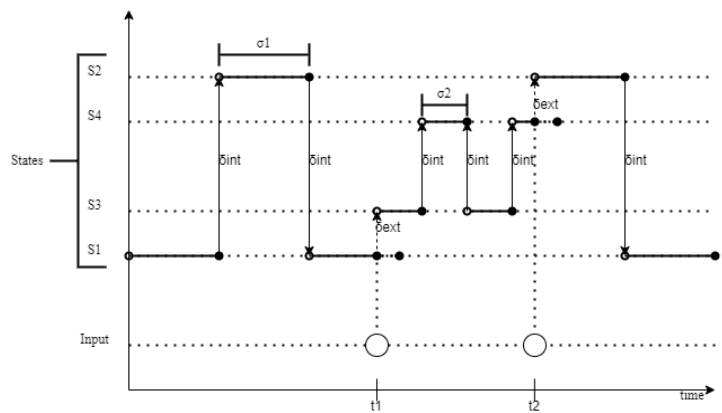

Figure 4: Blinky state diagram

The blinky atomic has a set of 4 states S = {S1, S2, S3, S4}. The internal transition function transitions Blinky from state S1 to S2 and vice versa after a period of σ1 time units. Similarly, Blinky transitions from S3 to S4 and vice versa after a period of σ2 time units. Formally, $S1 = \delta_{int}(S2), S2 = \delta_{int}(S1); S3 = \delta_{int}(S4), S4 = \delta_{int}(S3)$. The external transition function (that occurs when an external input is detected) transitions the Blinky from S1 to S3 and vice versa or from S2 to S4 or vice versa depending on the present state. Formally, $S2 = \delta_{ext}(S4), S4 = \delta_{ext}(S2); S1 = \delta_{ext}(S3), S3 = \delta_{ext}(S1)$. Upon observing Figure 4, the Blinky system toggles between S1 and S2 with a period of σ1 until time t1. At time t1, and external input is received, which triggers $\delta_{ext}(S1)$, transitioning the system from S1 to S3. The system then continues oscillation between S3 and S4 with a period of σ2 until time t2. At time t2, another input is received, which again triggers the external transition function $\delta_{ext}(S4)$, transitioning the system from S4 to S2. Every internal transition function calls the output function $\lambda(S1), \lambda(S2), \lambda(S3), \lambda(S4)$ respectively depending on the present state (state prior to transition).

The complete system comes together as the BlinkySystem coupled model which is the top model that integrates all the atomic models together.

**RESULTS AND DISCUSSION**

The system was implemented on the ESP32 using the adopted RT-CADMIUM libraries. The simulation output can be observed in Table 1 and the logging output received from the ESP32 is shown in Figure 5 and Table 2

*Table 1: Blinky simulation output*

| time | model_id | model_name | port_name | data |
|---|---|---|---|---|
| 4 | 1 | blinky | out | 1 |
| 4.5 | 1 | blinky | out | 0 |
| 5 | 1 | blinky | out | 1 |
| … | … | …. | … | … |
| 28.5947 | 2 | generator | out | 0 |
| 29.5947 | 1 | blinky | out | 0 |
| 30.5947 | 1 | blinky | out | 1 |
| 31.5947 | 1 | blinky | out | 0 |

The system was implemented on the ESP32 using the adopted RT-CADMIUM libraries. The simulation output can be observed in Table 1 and the logging output received from the ESP32 is shown in Figure 5 and Table 2

Table 1 shows the logger output when simulating the BlinkySystem model. The logger shows the values of the present at the output ports of every model along with a timestamp and the unique model id. The characteristics of the BlinkySystem can be observed from the simulation output and can be cross verified against the state diagram in Figure 4. As the diagram suggests, the system can be seen to be oscillating between state output 1 and state output 0 with a period of 0.5s. The first three rows of the table portray this trend. At around the 29.6$^{th}$ second, the generator with model ID 2 is seen to produce an output, which feeds into the Blinky input. According to the state diagram (Figure 4), an external input triggers a change in the state oscillation frequency. The same can be observed in Table 1, when the generator produces an output at the 28.5947$^{th}$ second, the period (σ) changes from 0.5s to 1s.

The implementation of the same on the ESP32 also produces a logging output portrayed in Figure 5.

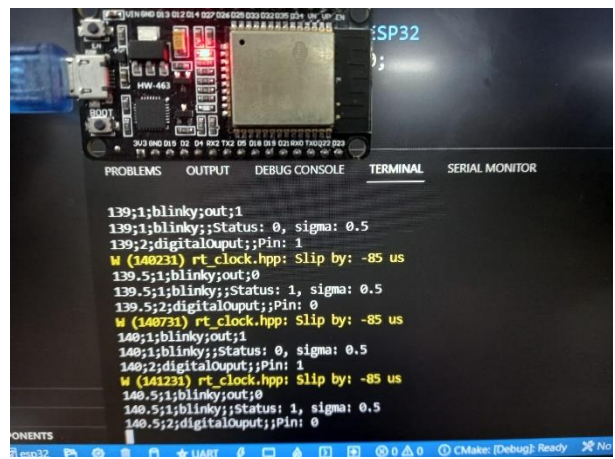

*Figure 5: ESP32 output*

Figure 5 Shows the output of the ESP32 executing the Blink program. This output serial stream can be parsed into a csv file as shown in Table 2

*Table 2: Formatted output of the ESP32*

| time | model_id | model_name | port_name | data |
|---|---|---|---|---|
| 172 | 1 | blinky | out | 1 |
| 172 | 2 | digitalOuput | | Pin: 1 |
| 172.5 | 1 | blinky | out | 0 |
| 172.5 | 2 | digitalOuput | | Pin: 0 |
| 173 | 1 | blinky | out | 1 |
| 173 | 2 | digitalOuput | | Pin: 1 |
| 173.6 | 3 | digitalInput | out | 1 |
| 174.6 | 1 | blinky | out | 0 |
| 174.6 | 2 | digitalOuput | | Pin: 0 |
| 175.6 | 1 | blinky | out | 1 |

A closer look at Table 2 shows the similarity of the actual output to the simulation. Here, unlike the simulation, 2 models with model ID 1 and 2 seem to be oscillating initially with σ = 0.5s. the 2 models that are oscillating are Blinky and digitalOutput. During simulation, digitalOutput model remains disconnected and hence is absent in the simulation. Apart from the presence of digitalOutput, the logger output of the deployed system is comparable to the simulation output. At the 173.6$^{th}$ second the button press is registered, and the σ of oscillation changes from 0.5s to 1s.

## PERFORMANCE EVALUATION

Metrics like memory footprint and response time can be used as external parameters to measure the performance of the system. Further, the amount by which the scheduler has slipped (within the range defined by the MISSED_DEADLINE_TOLERANCE value) will also enable us to measure the performance of the system.

```
Total sizes:
Used static DRAM:   18216 bytes ( 162520 remain, 10.1% used)
      .data size:   10088 bytes
      .bss  size:    8128 bytes
Used static IRAM:   49226 bytes (  81846 remain, 37.6% used)
      .text size:   48199 bytes
    .vectors size:   1027 bytes
Used Flash size :  454003 bytes
         .text  :  321647 bytes
         .rodata:  132100 bytes
Total image size:  513317 bytes (.bin may be padded larger)
```

*Figure 6: Output memory map of idf.py build*

Figure 6 shows the image size of the binary after compilation of the Blinky source code. The parameter of focus for performance evaluation would be the flash size. Flash size describes the size of the binary image after compilation of source code. As per Figure 6 the binary size is shown to be 454003 bytes or 443.3 kB. This is however not the final size of the image loaded into the hardware.

```
* Executing task: C:\Users\appus\.espressif\python_env\idf5.0_py3.8_env\Scripts\python.exe d:\esp-idf\components\esptool_py\esptool\esptool.py -p COM
5 -b 460800 --before default_reset --after hard_reset --chip esp32 write_flash --flash_mode dio --flash_freq 40m --flash_size detect 0x1000 bootloader/
bootloader.bin 0x10000 main.bin 0x8000 partition_table/partition-table.bin

esptool.py v4.4
Serial port COM5
Connecting........................
Chip is ESP32-D0WDQ6 (revision v1.0)
Features: WiFi, BT, Dual Core, 240MHz, VRef calibration in efuse, Coding Scheme None
Crystal is 40MHz
MAC: 94:b5:55:1a:ab:a0
Uploading stub...
Running stub...
Stub running...
Changing baud rate to 460800
Changed.
Configuring flash size...
Auto-detected Flash size: 4MB
Flash will be erased from 0x00001000 to 0x00007fff...
Flash will be erased from 0x00010000 to 0x0008efff...
Flash will be erased from 0x00008000 to 0x00008fff...
Compressed 26336 bytes to 16382...
Wrote 26336 bytes (16382 compressed) at 0x00001000 in 0.9 seconds (effective 229.9 kbit/s)...
Hash of data verified.
Compressed 518304 bytes to 242067...
Wrote 518304 bytes (242067 compressed) at 0x00010000 in 6.5 seconds (effective 642.5 kbit/s)...
Hash of data verified.
Compressed 3072 bytes to 105...
Wrote 3072 bytes (105 compressed) at 0x00008000 in 0.1 seconds (effective 313.0 kbit/s)...
Hash of data verified.

Leaving...
Hard resetting via RTS pin...
```

*Figure 7: Output of the flash tool*

Figure 7 shows the output of the flash tool (idf.py flash). This output gives us a deeper insight into the size of the binary image. Here, we can observe that the image is written in three separate address spaces, namely: 0x00001000, 0x00010000 and 0x00008000. Referring to the ESP-IDF documentation[21] regarding the partition tables, we can observe that this partitioning is the default partitioning scheme for the build configuration of 'single factory app (large), no OTA'. Each section is offset by a block multiple of 4kB (0x1000). From Figure 7, we can observe that the largest data block is stored in address 0x00010000 (corresponding to an offset of 64kB). This is where the bootloader starts its execution. We can see that the size of the data loaded into this partition is 518304 bytes, this is the final binary size including the padding. But we can observe that this is then compressed to 242067 bytes or 236.4kB. So, in practice, the binary size can be said to be **242067 bytes**.

```
| Module                     |            .text |         .data |          .bss |
|----------------------------|------------------|---------------|---------------|
| [fill]                     |        146(+146) |         9(+9) |       27(+27) |
| [lib]/c.a                  |    72504(+72504) |  2574(+2574)  |       97(+97) |
| [lib]/gcc.a                |      7256(+7256) |         0(+0) |         0(+0) |
| [lib]/misc                 |        188(+188) |         4(+4) |       28(+28) |
| [lib]/nosys.a              |          32(+32) |         0(+0) |         0(+0) |
| [lib]/stdc++.a             |  175004(+175004) |    145(+145)  |   5720(+5720) |
| mbed-os/drivers            |        906(+906) |         0(+0) |         0(+0) |
| mbed-os/hal                |      1424(+1424) |         4(+4) |       66(+66) |
| mbed-os/platform           |      3006(+3006) |    260(+260)  |     225(+225) |
| mbed-os/targets            |      5806(+5806) |         4(+4) |     348(+348) |
| top_model/main_rt_model.o  |    15196(+15196) |         0(+0) |         1(+1) |
| Subtotals                  |  281468(+281468) |  3000(+3000)  |   6512(+6512) |
Total Static RAM memory (data + bss): 9512(+9512) bytes
Total Flash memory (text + data): 284468(+284468) bytes
```

*Figure 8: Output of mbed compile (image by Ezequiel Pecker-Marcosig[20])*

Figure 8 shows the compile output when the same Blinky system is compiled using the Mbed OS version of CADMIUM[20]. Here, we can observe that the total binary size comes upto 284468 bytes. This is 41.4 kB more that the ESP-IDF implementation of the same. Considering the image sizes are 200+ kB, a difference of 42 kB may not seem significant, but, considering the available storage size, every byte counts. Fig, 10 shows the output when Blinky is compiled for the STM32 Nucleo-F401RE that has a total of 512kB of flash storage. Hence, to take the flash storage into consideration, the percentage consumption of both the binaries gives a better overview of memory footprint.

For a performance benchmark, the blinky was also compiled without CADMIUM. Figure 9 shows the build and flash outputs.

```
Total sizes:
Used static DRAM:   12096 bytes ( 168640 remain, 6.7% used)
      .data size:    9816 bytes
      .bss  size:    2280 bytes
Used static IRAM:   47414 bytes (  83658 remain, 36.2% used)
      .text size:   46387 bytes
   .vectors size:    1027 bytes
Used Flash size :  127847 bytes
      .text   :     92435 bytes
      .rodata :     35156 bytes
Total image size:  185077 bytes (.bin may be padded larger)
```

```
Configuring flash size...
Flash will be erased from 0x00001000 to 0x00007fff...
Flash will be erased from 0x00010000 to 0x0003dfff...
Flash will be erased from 0x00008000 to 0x00008fff...
Compressed 26368 bytes to 16417...
Wrote 26368 bytes (16417 compressed) at 0x00001000 in 0.9 seconds (effective 241.3 kbit/s)...
Hash of data verified.
Compressed 185216 bytes to 96517...
Wrote 185216 bytes (96517 compressed) at 0x00010000 in 2.9 seconds (effective 517.7 kbit/s)...
Hash of data verified.
Compressed 3072 bytes to 103...
Wrote 3072 bytes (103 compressed) at 0x00008000 in 0.1 seconds (effective 305.4 kbit/s)...
Hash of data verified.

Leaving...
Hard resetting via RTS pin...
```

*Figure 9(a)(b): Output memory map of idf.py build (Blinky sans CADMIUM) and the flash output (from left to right)*

From Figure 9(a) we can observe that the boiler plate blinky code alone takes up 185077 bytes of storage. After compression, this is reduced to 96517 bytes (as seen in Figure 9(b)). Hence, CADMIUM implemented directly on ESP-IDF takes up 142kB of memory.

The ESP32 having a total of 4MB of flash, the percentage consumption comes to:

- 4.4% before compression (185077 bytes, without CADMIUM)
- 2.3% after compression (97517 bytes, without CADMIUM)
- 12.3% before compression (518304 bytes, with CADMIUM)
- 5.7% after compression (242067 bytes, with CADMIUM)

The Mbed OS version on the Nucleo board with 512 kB of flash storage comes to:

- 54.2% no compression algorithm is run prior to flashing.

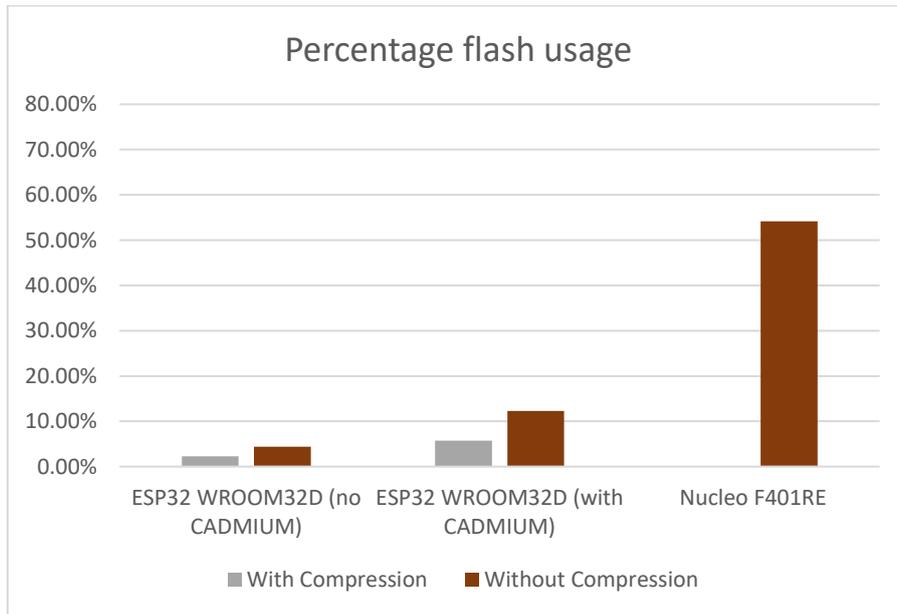

*Figure 10: Chart showing percentage consumption of flash vs embedded boards.*

Figure 11 shows the simulation output of BlinkySystem on the Nucleo F401RE. It is similar to the output seen in Figure 5. Upon observation of similarities, the 'deadline slip' values stand out. On the ESP32 implementation, the deadline slips by 85 microseconds, that is, the scheduler overshoots the deadline time by 85 microseconds. On the other hand, the STM output shows that the deadline is missed be 85,629 microseconds. This is an improvement of more than 10000%.

*Figure 11: STM logger output*

## CONCLUSION

DEMES methodology brings the rigour of formal specification to practical systems implemented on embedded platforms. The RT-CADMIUM platform allowed users to bring their models into the physical world by following the DEVS formalism, so long as they used platforms compliant with Mbed OS. This shortcoming of the ecosystem was identified.

This issue was tackled dichotomously. The foundation of RT-CADMIUM was changed to the newer version of CADMIUM[7], and Mbed OS was removed in favour of bare-metal implementation. A case study was conducted to evaluate the performance of the system. The Blinky system was deployed on the adapted version of RT-CADMIUM, and the previous Mbed OS version of RT-CADMIUM. The performance was measured based on two metrics; memory footprint of the source image on the device flash, and the time taken to execute the transition functions. The memory footprint didn't see a drastic change between the versions of RT-CADMIUM, however, considering the percentage consumption, newer version was found to be drastically more space efficient. Similarly, the time taken to execute transitions was observed to be much lesser than the previous version of RT-CADMIUM. These improvements in execution time and memory footprint shows that this development methodology is growing closer to commercialization.

## REFERENCES

[1]  B. Earle, K. Bjornson, C. Ruiz-Martin, and G. Wainer, "Development of A Real-Time Devs Kernel: RT-Cadmium," *Proceedings of the 2020 Spring Simulation Conference, SpringSim 2020*, May 2020, doi: 10.22360/SPRINGSIM.2020.CPS.002.

[2]  S. M. A. Group *et al.*, "Internet of Things (IoT): A Literature Review," *Journal of Computer and Communications*, vol. 03, no. 05, pp. 164–173, 2015, doi: 10.4236/JCC.2015.35021.

[3]  L. Belloli, D. Vicino, C. Ruiz-Martin, and G. Wainer, "BUILDING DEVS MODELS WITH THE CADMIUM TOOL".

[4]  V. Kumar, "Real-Time Scheduling Algorithms".

[5]  G. Wainer Joseph Boi-Ukeme, "Applying Modelling and Simulation for Development of Embedded Systems".

[6]  A. C. H. Chow and B. P. Zeigler, "Parallel DEVS: a parallel, hierarchical, modular modeling formalism," *Winter Simulation Conference Proceedings*, pp. 716–722, 1994, doi: 10.1109/WSC.1994.717419.

[7]  R. Cárdenas and G. Wainer, "Asymmetric Cell-DEVS models with the Cadmium simulator," *Simul Model Pract Theory*, vol. 121, Dec. 2022, doi: 10.1016/J.SIMPAT.2022.102649.

[8]   J. S. Hong, T. Gon, K. Kyu, and H. O. Park, "A Real-Time Discrete Event System Specification Formalism for Seamless Real-Time Software Development," vol. 7, pp. 355–375, 1997.

[9]   B. Zeigler and A. Muzy, "Theory of Modeling and Simulation Discrete Event and Iterative System Computational Foundations," 2019, Accessed: Apr. 04, 2023. [Online]. Available: https://www.elsevier.com/books-and-journals

[10]  "Development boards | Mbed." https://os.mbed.com/platforms/ (accessed Apr. 04, 2023).

[11]  S. Yoo and A. A. Jerraya, "INTRODUCTION TO HARDWARE ABSTRACTION LAYERS FOR SOC," 2003.

[12]  "Simple and efficient software development with the SimpleLink^TM MCU platform." https://www.ti.com/lit/wp/swsy004e/swsy004e.pdf?ts=1680690846988&ref_url=https%253A%252F%252Fwww.google.com%252F (accessed Apr. 05, 2023).

[13]  "Description of STM32F4 HAL and low-layer drivers." https://www.st.com/resource/en/user_manual/um1725-description-of-stm32f4-hal-and-lowlayer-drivers-stmicroelectronics.pdf (accessed Apr. 05, 2023).

[14]  "Hardware Abstraction Layer," 2017.

[15]  N. Semiconductor and F. Semiconductor, "Software support for microcontollers featuring the ARM ® Cortex ®-M core", Accessed: Apr. 05, 2023. [Online]. Available: www.twitter.com/nxp

[16]  S. Rounaq and M. Iqbal, "Vision, Challenges and Future Perspectives of Low Constrained Devices IOT Operating Systems: A Systematic Mapping Review," *European Journal of Engineering and Technology Research*, vol. 5, no. 12, pp. 107–115, Dec. 2020, doi: 10.24018/EJENG.2020.5.12.2284.

[17]  "ESP32WROOM32D & ESP32WROOM32U Datasheet," 2023, Accessed: Apr. 05, 2023. [Online]. Available: https://www.espressif.com/en/support/download/documents.

[18]  M. Babiuch, P. Foltýnek, and P. Smutný, "Using the ESP32 Microcontroller for Data Processing; Using the ESP32 Microcontroller for Data Processing," *2019 20th International Carpathian Control Conference (ICCC)*, 2019, doi: 10.1109/CarpathianCC.2019.8765944.

[19]  A. Maier, A. Sharp, and Y. Vagapov, "Comparative analysis and practical implementation of the ESP32 microcontroller module for the internet of things," *2017 Internet Technologies and Applications, ITA 2017 - Proceedings of the 7th International Conference*, pp. 143–148, Nov. 2017, doi: 10.1109/ITECHA.2017.8101926.

[20]  "jupyter-cadmiumv2/cadmiumv2_logging_howto.ipynb at master · epecker/jupyter-cadmiumv2." https://github.com/epecker/jupyter-cadmiumv2/blob/master/cadmiumv2_logging_howto.ipynb (accessed Apr. 10, 2023).

[21]  "Partition Tables - ESP32 - — ESP-IDF Programming Guide latest documentation." https://docs.espressif.com/projects/esp-idf/en/latest/esp32/api-guides/partition-tables.html (accessed Apr. 10, 2023).